\documentclass[12pt]{article}
\usepackage{amssymb,amsmath}
\usepackage[noblocks]{authblk}
\usepackage[top=0.75in, bottom=0.75in, left=0.75in, right=0.75in, dvips]{geometry}
\usepackage{caption}  
\date{}   

\begin{document}
\textwidth 10.0in       
\textheight 9.0in 
\topmargin -0.60in
\title{The Superparticle on the Surface $S_2$}
\author[]{D.G.C. McKeon\thanks{Email: dgmckeo2@uwo.ca}}
\affil[] {Department of Applied Mathematics, The
University of Western Ontario, London, ON N6A 5B7, Canada}
\affil[] {Department of Mathematics and
Computer Science, Algoma University, Sault St.Marie, ON P6A
2G4, Canada}
\maketitle

\maketitle

\begin{abstract}
A superparticle action which is globally supersymmetric in the target space is proposed.  The supersymmetry is the supersymmetric extension of the rotation group $O(3)$.
\end{abstract}

\section{Introduction}
The superparticle action [1,2] was designed to be invariant under a supersymmetry transformation in the target space.  This supersymmetry was taken to be the supersymmetric extension of the Poincar$\acute{e}$ group.

It is possible to make supersymmetric extensions of other groups; in particular groups of transformations on spaces of constant curvature [3].  In this paper we devise an action for a superparticle which has a target space invariance that is a supersymmetric extension of the rotation group $O(3)$, associated with the sphere $S_2$ defined by
\begin{equation}
(x^1)^2 + (x^2)^2 + (x^3)^2 = r^2.
\end{equation}

\section{The Superparticle}

The rotation group $O(3)$ leaves the quadratic form of eq. (1) invariant.  The generators of this group, $J^a$ have the algebra defined by 
\begin{equation}
[J^a, J^b] = i\epsilon^{abc} J^c.
\end{equation}
One can extend this group [3,4] by introducing Fermionic spinor generators $Q_i, Q_i^\dagger$ plus an additional Bosonic generator $Z$ so that one has the algebras defined by the (anti-)commutators
\begin{subequations}
\begin{align}
\left\lbrace Q_i, Q_j^\dagger \right\rbrace &= Z \delta_{ij} \mp 2 \tau_{ij}^a J^a \\
[J^a, Q_i ] &= -\frac{1}{2} (\tau^a Q)_i \\
[Z, Q_i] &= \mp Q_i
\end{align}
\end{subequations}
in addition to eq. (2).  The Pauli matrices $\tau^a$ satisfy
\begin{subequations}
\begin{align}
\tau^a \tau^b &= \delta^{ab} + i \epsilon^{abc} \tau^c\\
\tau^a_{ij} \tau^a_{k\ell} &= 2\delta_{i\ell}\delta_{kj} - \delta_{ij} \delta_{k\ell}\\
\tau_{ij}^a \delta_{k\ell} + \tau_{k\ell}^a \delta_{ij} &= \tau_{i\ell}^a \delta_{kj} + \tau_{kj}^a \delta_{i\ell}\\
\epsilon^{abc} \tau_{ij}^b \tau_{k\ell}^c &= i\left( \tau_{i\ell}^a \delta_{kj} - \tau_{kj}^a \delta_{i\ell} \right) \\
\tau_{ij}^a \tau_{k\ell}^b &= \frac{1}{2}\left[ \delta^{ab} \delta_{i\ell}\delta_{kj} + i \epsilon^{abc} \left( \tau_{i\ell}^c \delta_{kj} - \delta_{i\ell}\tau_{kj}^c\right) + \tau_{i\ell}^a \tau_{kj}^b + \tau^b_{i\ell} \tau_{kj}^a - \delta^{ab} \tau_{i\ell}^c \tau_{kj}^c \right].
\end{align}
\end{subequations}
(There is an additional extension of the $O(3)$ algebra whose (anti)-commutators are
\[\left\lbrace Q_i, \tilde{Q}_j \right\rbrace = \tau_{ij}^a J^a \quad  
\left\lbrace Q_i, Q_j^\dagger \right\rbrace = \tau_{ij}^a Z^a \eqno(5a,b) \]
\[ [J^a, Q_i ] = -\frac{1}{2} \tau_{ij}^a Q_j \quad [Z^a, \tilde{Q}_i] = \frac{1}{2} Q^\dagger_j \tau_{ji}^a \eqno(5c,d)\]
\[ [J^a, J^b ] = i\epsilon^{abc} J^c \quad [Z^a, Z^b] = -i \epsilon^{abc} J^c \eqno(5e,f) \]
\[ [J^a, Z^b ]= i \epsilon^{abc} Z^c \eqno(5g) \]
where $\tilde{Q} = Q^T \tau^2$ and $Z^a$ is a Bosonic vector.)

We now introduce Bosonic vector coordinates $x^a(\tau)$, a Bosonic scalar coordinate $\beta(\tau)$, and Fermionic spinor coordinates $\theta_i(\tau)$ for a superparticle moving along a trajectory parameterized by $\tau$ on the sphere $S_2$ defined by eq. (1).  Next, we define
\[ Y^a = \dot{\beta} x^a - \beta \dot{x}^a + \epsilon^{abc} x^b \dot{x}^c + i\left( \dot{\theta}^\dagger \tau^a\theta - \theta^\dagger \tau^a \dot{\theta}\right)\eqno(6) \]
and note that $\delta Y^a = 0$ under the supersymmetry transformation
\[ \delta\beta = -\epsilon^\dagger \theta - \theta^\dagger \epsilon \eqno(7a)\]
\[ \delta \theta = (-i \tau \cdot x - \beta )\epsilon \eqno(7b) \]
\[ \delta \theta^\dagger = \epsilon^\dagger (i \tau \cdot x - \beta )\eqno(7c) \]
\[ \delta x^a = i (\epsilon^\dagger \tau^a \theta - \theta^\dagger \tau^a \epsilon ) \eqno(7d) \]
where $\epsilon$ is a constant Fermionic Dirac Spinor.  In addition, the quantity
\[R^2 = x^a x^a - 2\theta^\dagger \theta + \beta^2 \eqno(8)\]
is also invariant under the transformation of eq. (7).  For an ordinary massless particle moving on a sphere $S_2$ we would have the Lagrangian
\[ L_0 = \frac{1}{2e} \left(\epsilon^{abc} x^b \dot{x}^c\right)^2 + \lambda \left( x^a x^a - r^2 \right) \eqno(9) \]
where $\lambda(\tau)$ is a Lagrange multiplier and $e(\tau)$ is an einbein field.  This we now generalize to 
\[ L = \frac{1}{2e} Y^aY^a + \lambda \left( x^ax^a - 2\theta^\dagger \theta + \beta^2 - R^2\right).\eqno(10) \]
This Lagrangian is obviously invariant under the global supersymmetry transformation of eq. (7).

The generator of the transformation of eq. (7) is $\epsilon^\dagger Q + Q^\dagger \epsilon$ where
\[ Q = - \theta \frac{\partial}{\partial\beta} + (i\tau \cdot x - \beta)  \frac{\partial}{\partial \theta^\dagger} + i \tau \cdot \nabla\theta \eqno(11a) \]
\[ Q^\dagger = - \theta^\dagger \frac{\partial}{\partial\beta} +
 \frac{\partial}{\partial \theta} (i\tau \cdot x + \beta)  - i \theta^\dagger\tau \cdot\nabla\;. \eqno(11b) \]
These generators appear also in refs. [3,4].  By using eq. (4), it can be verified that if 
\[ \delta_i A = \left( \epsilon_i^\dagger Q + Q^\dagger\epsilon_i\right)A \eqno(12) \]
then the Jacobi identity
\[ \big( \left[ \delta_1, \left[ \delta_2, \delta_3\right]\right] + 
\left[ \delta_2, \left[ \delta_3, \delta_1\right]\right] + 
\left[ \delta_3, \left[ \delta_1, \delta_2\right]\right]\big) A = 0 \eqno(13) \]
is satisfied.

The canonical momenta conjugate to $(x^a, \beta, \lambda, e, \theta^\dagger, \theta)$ respectively are given by 
\[ p^a = \frac{1}{e} \left( - \beta Y^a + \epsilon^{abc} Y^b x^c \right) \eqno(14a)\]
\[\hspace{-2.48cm} p_\beta = \frac{1}{e} x^aY^a \eqno(14b)  \]
\[\hspace{-3.3cm} p_\lambda = 0 \eqno(14c) \]
\[\hspace{-3.3cm} p_e = 0 \eqno(14d) \]
\[\hspace{-1.8cm} \pi = \frac{i}{e} \;\;(\tau \cdot Y \theta ) \eqno(14e) \]
\[ \hspace{-1.8cm}\pi^\dagger = \frac{i}{e} (\theta^\dagger \tau \cdot Y) .\eqno(14f) \]
(We use the left hand derivative for Fermionic variables.)  From eqs. (14a,b) it follows immediately that we have the primary constraint
\[ \Pi = x^a p^a + \beta p_\beta = 0. \eqno(15) \]
Furthermore, from eq. (14a) we obtain
\[ Y^a = \frac{e}{x^2+\beta^2} \left( -\beta\delta^{ab} + \epsilon^{apb} x^p - \frac{1}{\beta} x^ax^b\right) p^b \eqno(16) \]
and so eqs. (14d,e) result in two more primary constraints
\[\hspace{-.2cm} \chi = \pi - i\, \Xi\, \theta = 0 \eqno(17a) \]
\[ \chi^\dagger = \pi^\dagger - i  \theta^\dagger \Xi = 0 \eqno(17b) \]
where
\[ \Xi = \frac{1}{x^2+\beta^2} \left( -\beta p^a + \epsilon^{abc} x^b p^c - \frac{1}{\beta} x^a x \cdot p \right) \tau^a . \]
Eq. (17) is immediately seen to provide a pair of primary second class constraints as\footnote{For Poisson Brackets (PB) involving Bosonic variables $B_i$ and Fermionic variables $F_i$, depending on Bosonic canonical pairs $(q, p)$ and Fermionic canonical pairs $(\psi , \pi)$, we use the conventions
\[ \left\lbrace B_1, B_2\right\rbrace = \left( B_{1,q} B_{2,p} - B_{2,q} B_{1,p}\right) + \left( B_{1,\psi} B_{2,\pi} - B_{2,\psi} B_{1,\pi}\right)\nonumber \]
\[\hspace{.5cm}\left\lbrace B,F \right\rbrace = - \left\lbrace F, B \right\rbrace = \left( B_{,q} F_{,p} - F_{,q} B_{,p} \right) +
\left( B_{,\psi} F_{,\pi}+ F_{,\psi} B_{,\pi} \right) \nonumber \]
\[ \left\lbrace F_1, F_2 \right\rbrace = \left( F_{1,q} F_{2,p} + F_{2,q} F_{1,p} \right) - \left( F_{1,\psi} F_{2,\pi} + F_{2,\psi} F_{1,\pi} \right).\nonumber \]} (using eq. (4e))
\[ \left\lbrace \chi , \chi^\dagger \right\rbrace = \frac{2i}{\Lambda\beta} \left( p \cdot \tau \theta^\dagger \theta - \theta^\dagger p \cdot \tau \theta \right) \eqno(18) \]
\[ \hspace{3cm}
+ \frac{4i}{\Lambda} \left( \Xi \theta^\dagger \theta - \theta^\dagger \Xi \theta \right) + 2i \Xi \qquad \left( \Lambda = x^2 + \beta^2\right)\begin{scriptsize}
{\footnotesize •\begin{small}
{\normalsize •}
\end{small}}
\end{scriptsize}.
\nonumber \]

The canonical Hamiltonian for our system is given by 
\[ H_c = \dot{\theta}^\dagger\pi + \dot{\theta}^T \pi^{\dagger T} + \dot{x}^ap^a + \dot{\beta} p_\beta - L \nonumber \]
\[\hspace{1.1 cm}= \frac{1}{2e} Y^aY^a - \lambda \left( x^2 - 2 \theta^\dagger \theta + \beta^2 - R^2 \right) \eqno(19) \]
which by eq. (16) becomes
\[ = \frac{1}{2} \frac{e}{x^2+\beta^2} (p^2 + p_\beta^2) - \lambda (x^2 - 2\theta^\dagger\theta + \beta^2 - R^2) \;.\eqno(20) \]
Eqs. (14c,d; 15; 17a,b) are all primary; eqs. (14c,d) imply the secondary constraints
\[ \Sigma_1 = x^2 - 2\theta^\dagger\theta + \beta^2 - R^2 = 0 \eqno(21a) \]
and
\[ \Sigma_2 = \frac{p^2 + p_\beta^2}{x^2 +  \beta^2} = 0. \eqno(21b) \]
The PB $\left\lbrace \Pi , \Sigma_2 \right\rbrace$ weakly vanishes but the PB $\left\lbrace \Pi , \Sigma_1 \right\rbrace$ does not.  We note that by eq. (18), all Fermionic constraints are second class; since there are no first class Fermionic constraints, there is no local Fermionic symmetry that would be the analogue of the $\kappa$-symmetry discussed in ref. [5].  

\section*{Conclusions}

We have introduced a Lagrangian that is invariant under a global supersymmetry transformation that is an extension of the rotation group $O(3)$.  

We note that the stereographic projection
\[ x^a = \frac{2\eta^ar^2}{r^2 + \eta^2} \quad (a=1,2) \eqno(22a) \]
\[\hspace{-1.2cm} x^3 = r\left( \frac{r^2 - \eta^2}{r^2 + \eta^2} \right) \eqno(22b) \]
can be used to map coordinates on the surface of a sphere $S_2$ defined by eq. (1) onto the Euclidean plane defined by $(\eta^1, \eta^2)$.  If instead of eq. (1), we have the constraint of eq. (21a), we would replace $r^2$ in eq. (22) with $R^2 + 2\theta^\dagger\theta - \beta^2$.

It would be of interest to find superparticle Lagrangians that are invariant under supersymmetric extensions of other symmetry groups associated with spaces of constant curvature, such as $AdS$ and $dS$ spaces. Quantization of such models would be non-trivial.

We note that there recently has been some interest in exploring supersymmetry in curved spaces; see for example ref. [9].

\section*{Acknowledgements}

We wish to thank Alex Buchel for conversations and R. Macleod for a timely suggestion.

\end{document}